\documentclass[aps,prl,epsf,superscriptaddress,amsmath,amssymb,amsfonts,twocolumn,showpacs,nofootinbib]{revtex4-1}

\usepackage[sectionbib]{bibunits}
\usepackage{bibunits}

\usepackage{graphicx}
\usepackage{dcolumn}
\usepackage{bm}
\usepackage{braket}
\usepackage{amsmath}
\usepackage{mathtools}
\usepackage{graphicx,color,xcolor}
\usepackage{hyperref}
\usepackage[capitalize]{cleveref}
\newcommand{\abs}[1]{\left| #1 \right|} 
\newcommand{\norm}[1]{\left\lVert#1\right\rVert}
\usepackage{multirow}

\usepackage[normalem]{ulem} 

\begin{document}

\begin{bibunit}[apsrev4-1]
    
\title{Generation of wave turbulence in dipolar gases driven across their phase transitions}

\author{George A. Bougas}
\email{gbougas@mst.edu}
\affiliation{Department of Physics and LAMOR, Missouri University of Science and Technology, Rolla, MO 65409, USA}

\author{Koushik Mukherjee}
\affiliation{Mathematical Physics and NanoLund, LTH, Lund University, Box 118, 22100 Lund, Sweden}%

\author{Simeon I. Mistakidis}
\affiliation{Department of Physics and LAMOR, Missouri University of Science and Technology, Rolla, MO 65409, USA}

\date{\today}

\begin{abstract}
Ultracold quantum gases with long-range anisotropic interactions host novel exotic phases of matter, such as supersolids, exhibiting both rigid and superfluid characteristics. 
The impact of this interplay on the out-of-equilibrium dynamics of dipolar gases, and in particular its connection with universal turbulent behavior, remains highly unexplored. 
Here, upon considering a dipolar Bose–Einstein condensate of dysprosium atoms being dynamically driven across the supersolid–superfluid phase transition and vice versa, we unveil the emergence of a robust nonequilibrium quasi-steady state. 
This state displays self-similar momentum distributions exhibiting algebraic decay at large momenta, with scaling exponents supporting the existence of wave turbulence. 
We demonstrate that supersolidity sustaining higher-lying momenta, associated with the roton minimum, promotes the development of turbulence. 
Our results provide a stepping stone toward unraveling and exploiting turbulent and self-similar behavior in anisotropically long-range interacting quantum gases amenable in current experiments.
\end{abstract}

\maketitle

{\vspace{0.25cm}\noindent\large\textbf{Introduction}} 
\label{sec:intro}

Turbulence is a ubiquitous hydrodynamic phenomenon associated with direct (inverse) energy cascades from larger (smaller) to smaller (larger) length scales~\cite{kolmogorov_local_1941,obukhov_distribution_1941,Kraichnan1967,frisch_turbulence_1995}.
It has an unambiguously interdisciplinary role determining the flow of fluids~\cite{hwang_airborne_2000,Cobelli_different_2011}, the formation of polymers~\cite{Choi_turbulent_2002}, the climate of planets~\cite{bakas_emergence_2013,Diamond_zonal_2005}, as well as star explosions~\cite{mosta_large_2015}.
Due to the many degrees of freedom involved, providing a microscopic description of turbulence is challenging. 
Instead, statistical models are capable to capture the energy transfer, mediated either through vortices or nonlinear waves, alias vortex and wave turbulence respectively~\cite{chorin_vortex_turbulence,yao_VT_2022, zakharov_kolmogorov_1992, nazarenko_wave_2011}.
The latter features turbulent cascades stemming from wave propagation in disparate media, such as classical and quantum fluids, optical systems, and plasma~\cite{nazarenko_wave_2011,dyachenko_optical_1992}. The energy transfer to smaller length scales occurs due to the nonlinear wave mixing~\cite{nazarenko_wave_2011,zakharov_kolmogorov_1992}. Importantly, a perturbative treatment of the nonlinear interactions leads to a unified statistical treatment, dubbed weak wave turbulence.

These complex transfer processes can be monitored in a controllable fashion in the realm of ultracold quantum gases~\cite{madeira_quantum_2020,tsatsos_quantum_2016,navon_emergence_2016,vinen_quantum_2002,Seman_route_2011,Yukalov_trapped_2023,Kwon_relaxation_2014}.
These platforms provide leverage on turbulence, allowing to track vortex reconnections and annihilations~\cite{Henn_turbulence_2009,Neely_characteristics_2013,gauthier_2019_giant,johnstone_2019_evolution,Baggaley_spectrum_2011,Mossman_dissipative_2018, Ghosh_turbulence,Das_2023,Bulgac_quantum_2024}, as well as the propagation of cascade fronts~\cite{Galka_emergence_2022,navon_synthetic_2019} and eventually the emergence of nonequilibrium steady states~\cite{dogra_universal_2023,martirosyan_equation_2024,zhu_turbulence_2024}.
Strikingly, initiation of inverse cascades in superfluids (SF)~\cite{karailiev_observation_2024}, and the formation of stratified turbulence by tuning the polarizability of dipolar cold atoms~\cite{bland_quantum_2018} have been demonstrated.

Long-range interacting gases~\cite{chomaz_dipolar_2022,lahaye_physics_2009,Hughes_accuracy_2023}, such as dipolar settings hosting exotic phases of matter, provide fertile platforms to control turbulent flow~\cite{sabari_vortex_2024,prasad_crow_2024}, as was already evinced in their classical counterparts, the ferrofluids~\cite{Altmeyer_transition_2015,Mouraya_stationary_2024,Altmeyer_magnetic_2015}.
These magnetic atomic systems commonly consisting of lanthanides~\cite{lu2011strongly, aikawa2012bose, miyazawa2022bose} sustain self-bound droplet arrays~\cite{kadau_observing_2016, schmitt_self-bound_2016, ferrier-barbut_observation_2016, chomaz_quantum-fluctuation-driven_2016, santos2016fluctuations, Wachtler_quantum_2016}, which can be even connected by a SF background forming supersolids (SS)~\cite{Chester1970, leggett_can_1970}.
Such states were observed in both elongated~\cite{tanzi2019supersolid, bottcher2019transient, chomaz2019long, tanzi2019supersolid} and planar geometries~\cite{Norcia2021a, Biagoni_dimensional_2022, Bland2022a}.
They originate from the softening of a roton minimum, modifying the momentum distribution with respect to that of a SF~\cite{petter_probing_2019, Guo2019}, which can be utilized to engineer the onset and characteristics of turbulence. In fact, the impact of such exotic phases of matter, featuring shear modulus and frictionless mass transport~\cite{Boninsegni2012,poli_excitations_2024},
on the turbulent response remains elusive. 
Moreover, since they require the presence of quantum fluctuations~\cite{lee1957eigenvalues,Zhang_phonon_2022,lima2011quantum} for their stabilization,  they share the premise to bring forth beyond mean-field aspects of turbulence as well.

\begin{figure*}[t!]
\centering
\includegraphics[width=\textwidth]{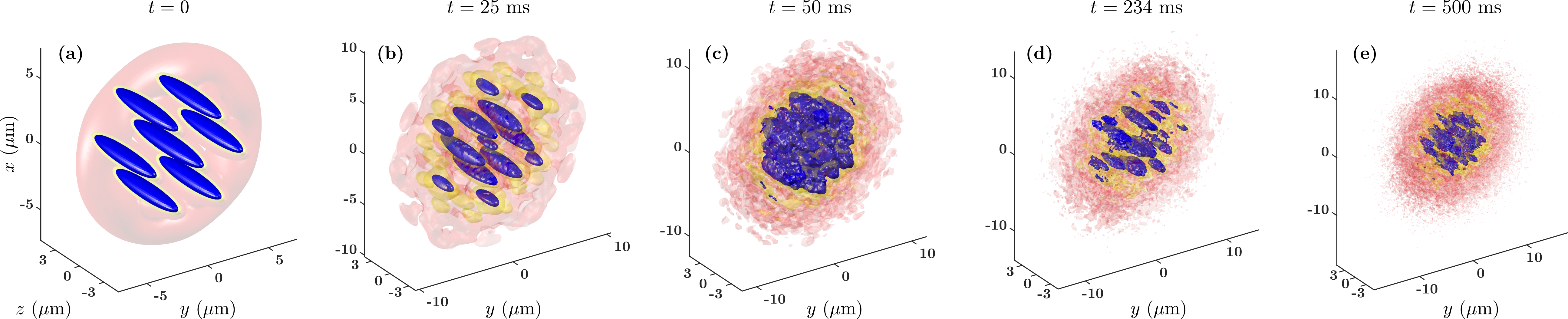}
\caption{
\textbf{Emergence of wave turbulence in a SS upon dynamically crossing the SS-to-SF  phase transition.}
(a)-(e) Three dimensional density isosurfaces [referring to $1\%$ (red), $10\%$ (yellow) and $25 \%$ (blue) of the 3D peak densities] at different time instants. 
The initial (final) 3D scattering length is $a_{\text{i}}=89 \, a_0$ ($a_{\text{f}}=98 \, a_0$) and the driving frequency $\omega_{\rm d}= 2\pi \times 127 \, \rm{Hz}$.
}
\label{Fig:Density_profiles}
\end{figure*}

In this theoretical work, we showcase the progressive emergence of quantum turbulence in the nonequilibrium dynamics of anisotropically long-range interacting 
dipolar Bose-Einstein condensates 
with the first-order quantum correction term.  
The dipolar gases are trapped in a cylindrically symmetric trap, leading to a 2D crystal arrangement in the SS phase.
The gases are initialized in
the SS or SF phases,
dynamically crossing the ensuing phase boundaries known from the ground state~\cite{Baillie_droplet_2018} by means of a modulated 3D scattering length, leading to constant energy injection. 
The turbulent response is captured by the self-similar long-time behavior of the momentum distribution 
at large wavenumbers, 
exhibiting an algebraic reduction with averaged scaling exponents 
$\gamma\approx 2.60$ and $\beta \approx - 0.63$, alluding to wave turbulence~\cite{Galka_emergence_2022,navon_emergence_2016}. 
These exponents allow us to characterize the emergent quasi-steady state~\cite{Galka_emergence_2022} in the momentum density distribution and were shown to provide an experimentally feasible way to extract a universal equation of state of weak wave turbulence~\cite{dogra_universal_2023}.
We also find that the long-time evolution of dipolar gases driven across their phase transitions follows the same universal turbulent description as many other systems.
The dynamics is characterized by a direct energy cascade front which becomes gradually statistically isotropic at the nonequilibrium quasi-steady state.   
The latter occurs irrespective of the initial state and the driving frequency when  dynamically crossing the phase transitions.
Interestingly, non-conventional 
SS and droplet states accelerate the manifestation of turbulence due to their extended momentum distribution.  
Finally, we further explicate that the above-described  phenomenology persists 
even under experimental conditions involving relevant three-body losses.

{\vspace{0.25cm}\noindent\large\textbf{Results}} \label{Sec:Results}

\textbf{Setup and driving protocol.} We consider a dipolar gas of $8\times 10^4$ $^{164}$Dy atoms polarized along the $z$-axis and confined in a cylindrically symmetric harmonic trap $V(\textbf{r})$ characterized by frequencies $(\omega_x,\omega_y,\omega_z)=2 \pi \times (43,43,133) \, \rm{Hz}$ 
similar to the experiments of Refs.~\cite{casotti_observation_2024,Bland2022a}. 
The magnetic atoms feature both short-range and long-range two-body interactions and their dynamics is modeled via the appropriate 3D  extended Gross-Pitaevskii equation~\cite{Wachtler_quantum_2016,Ronen_dipolar_2006,Yi_trapped_2001} (see also the Methods),    
\begin{gather}
\text{i} \hbar \partial_t \Psi(\textbf{r},t) = \Bigg [ - \frac{\hbar^2}{2m} \nabla^2 +V(\textbf{r})  + \frac{4\pi \hbar^2 a}{m} \abs{\Psi(\textbf{r},t)}^2  \nonumber \\ 
+ \int d\textbf{r}'~ U_{\text{dd}}(\textbf{r} -\textbf{r}')\abs{\Psi(\textbf{r}',t)}^2  
+  f (\epsilon_{\text{dd}}) \abs{\Psi(\textbf{r},t)}^3 \Bigg ] \Psi(\textbf{r},t).
\label{Eq:MF_equation}
\end{gather}
Here, $a$ refers to the 3D scattering length which is experimentally tunable via Fano-Feshbach resonances~\cite{Chin_Feshbach_2010,Maier_broad_2015}. Also, $m$ is the mass of $^{164}$Dy atoms, and $V(\textbf{r})$ designates the 3D harmonic trap.
The dipolar interaction $U_{\text{dd}}(\textbf{r}) = \frac{3\hbar^2 a_{\text{dd}}}{m}[\frac{1-3\cos^2\theta}{\textbf{r}^3}]$ contains the angle $\theta$ between the relative distance  $\textbf{r}=(x,y,z)$ of two dipoles 
and the $z$ quantization axis, while $a_{\text{dd}}=131 ~a_0$ is the $^{164}$Dy dipolar length, with 
$a_0$ 
being the Bohr radius. 
The fifth term is the repulsive first-order beyond mean-field Lee-Huang-Yang energy correction within the local density approximation, where $f(\epsilon_{\text{dd}})=\frac{128 \pi \hbar^2 a}{3m} \sqrt{a^3/\pi} (1+\frac{3}{2} \epsilon^2_{\text{dd}})$, 
and 
$\epsilon_{\text{dd}} \equiv a_{\text{dd}}/a$~\cite{lima2011quantum,schutzhold_mean_2006}.

It is known that in dipolar gases, the interplay between short-range and long-range interactions in the presence of quantum fluctuations yields a rich phase diagram~\cite{chomaz_dipolar_2022}. It features the  SF phase at $a/a_0>93$, as well as high-density exotic states being the SS for $87 \leq a/a_0 \leq 93$, and the droplet arrays at $a/a_0 \leq 86$ in our setup.

The dipolar system
is initialized in 
either a SF or a SS phase characterized by the 3D
scattering length $a_{\text{i}}$ (see Supplementary Note 5).
The scattering length
is subsequently periodically modulated with frequency $\omega_d$ according to $a(t)=a_{\text{i}}+(a_{\text{f}}-a_{\text{i}}) \sin^2(\omega_{\text{d}} t)$, where $\omega_{\text{d}} \gtrsim\omega_x$. 
Sufficiently large modulation amplitudes $(a_{\text{f}}-a_{\text{i}})$ trigger the dynamical crossing between different phases, arising at the ground state level at scattering lengths 
$a_{\text{i}}$ and $a_{\text{f}}$.

\textbf {Density profiles and direct energy cascade.} \label{sec:turbulence}
Initially, we fix
$a_{\text{i}}=89~a_0$ where a SS 
emerges. 
In quasi-2D geometries, such a state is typically characterized by spontaneously formed density peaks (droplets) arranged in a hexagonal structure 
on top of a SF background [Fig. \ref{Fig:Density_profiles}(a)]~\cite{Bland2022a,Baillie_droplet_2018}.

Setting $a_{\text{f}}=98~a_0$, 
representing a SF state at the ground state level, the dipolar gas is consecutively periodically 
driven across the SS-to-SF 
phase transition in a continuous manner. 
During the early dynamics  
the initial hexagonal crystal symmetry is gradually broken, as evidenced by the densities depicted in Fig.~\ref{Fig:Density_profiles}(b),(c), pertaining to $\omega_{\text{d}}/(2\pi) = 127 ~ \rm{Hz}$. 
This behavior is also reflected in the energy contribution of the first-order beyond mean-field correction term, which progressively becomes less important than the other energy terms as the crystals melt.
Instead, high density peaks start to  distribute randomly and the SF background becomes substantially perturbed. 
Meanwhile, a collective motion of the gas occurs, while a small number of defects in the form of vortex dipoles builds upon the SF background which annihilate or drift out of the condensate. 
The aforementioned density perturbation proliferates during 
the evolution 
[Fig. \ref{Fig:Density_profiles}(d),(e)], rendering the background distribution  
similar to that of a non-dipolar SF experiencing   
wave turbulence~\cite{Galka_emergence_2022,Zhu_direct_2023}.

\begin{figure}[t!]
\centering
\includegraphics[width=\columnwidth]{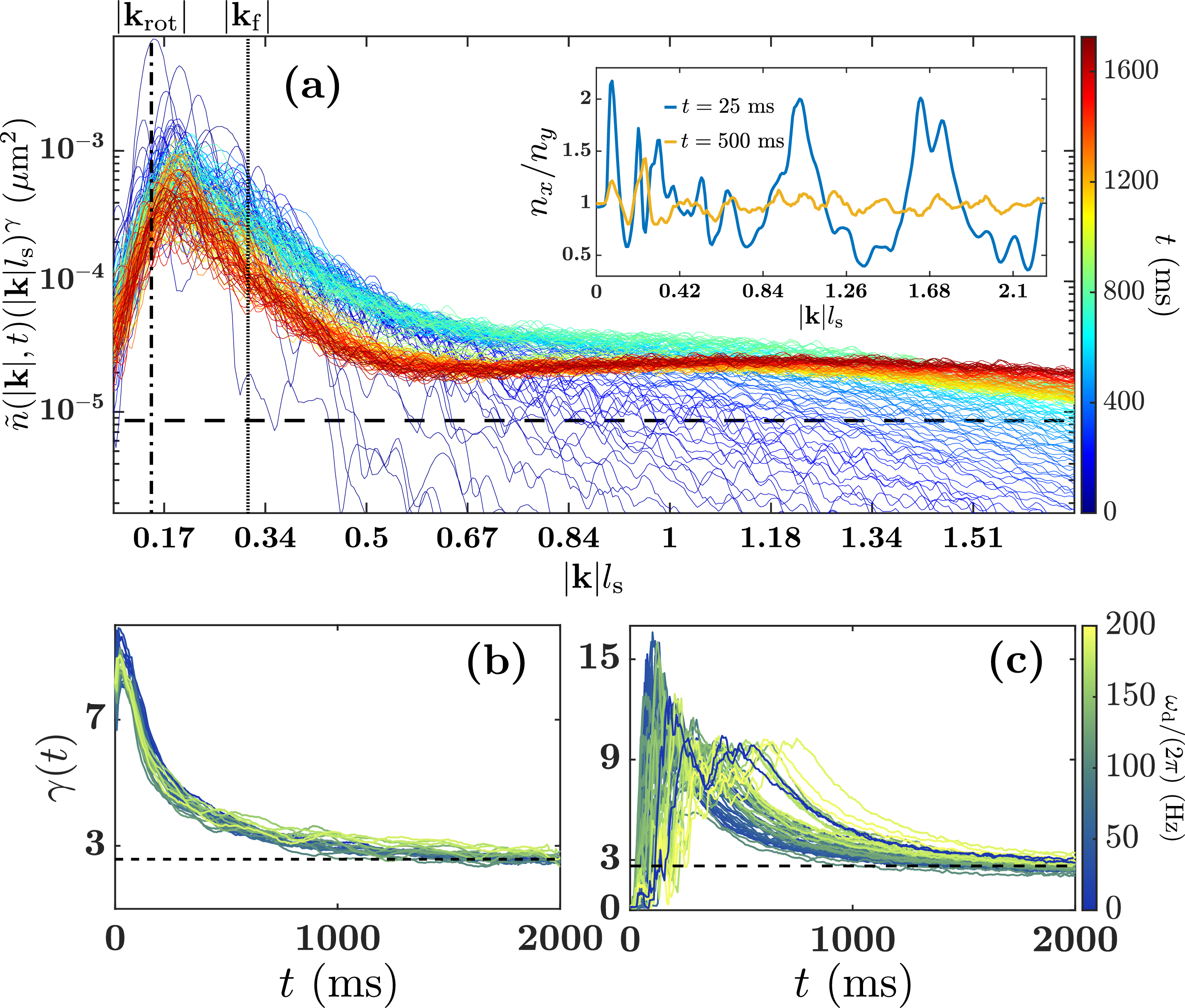}
\caption{
\textbf{Characterization of the self-similar turbulent response and its insensitivity on the driving frequency and initial state}. 
(a) Compensated momentum distribution $\tilde{n}(\abs{\textbf{k}},t) (\abs{\textbf{k}} l_{\rm s})^{\gamma}$ saturating to a plateau at large evolution times and momenta, obeying a power-law behavior with exponent $\gamma=2.5$. 
The dashed line determines the threshold for quantifying $\abs{\textbf{k}_{\rm cf}}$, 
and the vertical dash-dotted and dotted lines mark  $\abs{\textbf{k}_{\rm{rot}}}$ and the forcing wavenumber, $\abs{\textbf{k}_{\rm f}}$, respectively. 
The inset displays the ratio of integrated densities, $n_x/n_y$, signaling the isotropic dynamics at long evolution times. 
Panels (b) and (c) present the exponent dynamics $\gamma(t)$ with respect to the driving frequency, $\omega_{\rm d}$, in the case of the (b) SS-to-SF  transition  ($a_{\rm i}=89 \, a_0$, $a_{\rm f}=98 \, a_0$) and (c) vice versa ($a_{\rm i}=98 \, a_0$, $a_{\rm f}=91 \, a_0$).
The dashed lines represent the mean values of the exponents over all $\omega_{\rm d}$.
Other parameters are the same as in Fig.~\ref{Fig:Density_profiles}.}
\label{Fig:Momentum_exponents}
\end{figure}

The turbulent stage 
is independent of the driving frequency,
in contrast to the dynamical response 
at early times 
which strongly depends on both the driving characteristics and the initial state. 
To ascertain whether turbulence develops during the long time dynamics, we examine the momentum distribution  in the transverse plane defined as
$n(\textbf{k},t)= \abs{ \int d\textbf{r}_{\parallel}~ \text{e}^{-\text{i} \textbf{k}\cdot \textbf{r}_{\parallel}} \sqrt{ \int dz \abs{\Psi(\textbf{r}_{\parallel},z, t)}^2}  }^2 $, where $\textbf{k}=(k_x,k_y)$ and $\textbf{r}_{\parallel}=(x,y)$.
Since the most prominent effect of dipolar interactions is to induce a two-dimensional array of droplets characterized by a roton minimum, $k_{\text{rot}}$, lying in the two-dimensional plane, we perform our analysis focusing on the 2D momentum vector.
Probing this distribution, which can be experimentally monitored via 
line-of-sight absorption imaging~\cite{Galka_emergence_2022}, is motivated by 
the fact that the gas is significantly excited in the plane [see 
Fig.~\ref{Fig:Density_profiles}].
However, it holds that $\mu > \hbar \omega_z$, and therefore the transverse excitations across the $z$ direction are not completely frozen. They are rather mostly associated with the melting of the droplet crystals and the energy transfer (see Supplementary Note 1).
Strikingly, for $t>200 \, \rm{ms}$ 
$n(\textbf{k},t)$ becomes nearly isotropic 
at large $\abs{\textbf{k}}$~\cite{Galka_emergence_2022}, despite the presence of anisotropic interactions [see Eq. \eqref{Eq:MF_equation}], as evidenced by the ratio of integrated densities $n_x/n_y = \frac{\int dk_y \: n(\textbf{k},t)}{\int dk_x \: n(\textbf{k},t)}$ in the inset of Fig. \ref{Fig:Momentum_exponents}(a) for $\omega_{\text{d}}/(2\pi) = 127 ~ \rm{Hz}$.
Therefore the azimuthal average $\tilde{n}(\abs{\textbf{k}},t) = \int d\phi~ n(\abs{\textbf{k}},\phi,t)$ 
is 
the suitable measure for demonstrating turbulence properties~\cite{Neely_characteristics_2013}. 
The planar momentum distributions are normalized as $\int d\textbf{k} ~ n(\textbf{k},t)=1$ and $\int d\abs{\textbf{k}} ~ \abs{\textbf{k}} \tilde{n}(\abs{\textbf{k}},t)=1$.

The momentum distribution displays a power-law behavior at large momenta $\abs{\textbf{k}} \gg \abs{\textbf{k}_{\rm rot}}$, where $\abs{\textbf{k}_{\rm rot}} l_{\rm s} \simeq 0.1487 $ is the roton minimum. Here, $l_{\rm s} = \hbar / \sqrt{m \mu}$ is the length scale determined by the chemical potential $\mu$ of the initial state.
Fitting the $\tilde{n}(\abs{\textbf{k}},t)$ distribution with $\abs{\textbf{k}}^{-\gamma}$ in the region $\abs{\textbf{k}} \in [0.5,0.92]~ l_s^{-1}$ at $t=1700~\rm{ms}$ we extract the exponent $\gamma=2.50(04)$, 
alluding to wave turbulence~\cite{nazarenko_wave_2011},
which originates from nonlinear wave mixing~\cite{navon_emergence_2016}.
Note that the chosen momentum range for the fitting occurs at wavenumbers larger than the forcing wavenumber [vertical dotted line in Fig.~\ref{Fig:Momentum_exponents}(a)]. The latter is the momentum scale corresponding to the driving frequency, i.e. $\abs{\textbf{k}_{\rm f}} = 2\pi/\sqrt{\frac{\hbar}{m \omega_{\rm d}}}$, with $\omega_d/(2\pi) = 127 ~ \rm{Hz}$.

In the case of weak interactions 
where the generated waves possess random phases, the so-called weak wave turbulence regime is entered characterized by
$\gamma \simeq d$, where $d$ is the dimensionality~\cite{Kolmakov_wave_2014,Zhu_direct_2023,zakharov_kolmogorov_1992}.  
The discrepancy of $\gamma$ with the $d=2$ prediction stems from the fact that transfer of energy occurs also in the axial direction (see Supplementary Note 1).
Furthermore, a direct comparison of $\gamma$ with the $d=3$ value is hindered by the lack of isotropy of the 3D momentum distribution.
The latter therefore is expected to display a non-trivial momentum dependence, which cannot be easily inferred by the line-of-sight integrated $\tilde{n}(\abs{\textbf{k}},t)$ distribution and the associated $\gamma$ exponent.
Additional sources of deviation from the exponent estimates arise
due to the appearance of the small number of defects as also argued in the case of a non-dipolar SF undergoing wave turbulence~\cite{navon_emergence_2016}, higher-order corrections to the cascade scaling~\cite{Zhu_direct_2023}, long-range interactions, and the prominent in-plane distribution of the dipolar Bose-gas in the SS regime. 
The above-described turbulent response
holds for the non-viscous SS considered herein. 
However, the inclusion of viscosity through a phenomenological damping term diminishes 
short-wavelength excitations enforcing a rapid reduction of high-momentum tails and leading to larger values of $\gamma$ (see Supplementary Note 4).
Moreover, the addition of noise to the initial state leads to the same power-law behavior and exponent.

The momentum distribution compensated with $\abs{\textbf{k}}^{\gamma}$, evolves towards a plateau around $l_{\rm s}^{-1}$ [Fig. \ref{Fig:Momentum_exponents}(a)], 
demonstrating the development of a nonequilibrium quasi-steady state~\cite{dogra_universal_2023,navon_emergence_2016,Kolmakov_wave_2014}.
Before reaching that state, the slope of the compensated distribution's tail continuously decreases, marking the progression of a cascade front~\cite{navon_synthetic_2019} towards higher momenta. 
This is a clean manifestation  
of a direct energy cascade~\cite{Kolmakov_wave_2014}. 
Moreover, the amplitude of the high momentum peak associated with the roton minimum [vertical dashed line in Fig.~\ref{Fig:Momentum_exponents}(a)] decreases, a behavior directly stemming from the melting of the SS crystals.
The fitted exponent at these early times 
decreases before eventually approaching $\gamma \approx 2.5$. 
The power-law behavior is also inherited by the 2D kinetic energy density, $\mathcal{E}(\abs{\textbf{k}},t)$, near $\abs{\textbf{k}} l_{\rm s} \simeq 1$ as well, 
and the respective exponent saturates close to $-0.5$, as expected from the scaling relation, $\mathcal{E}(\abs{\textbf{k}},t) = \frac{\hbar^2 \abs{\textbf{k}}^2}{m} \tilde{n}(\abs{\textbf{k}},t)$.
All other interaction energy terms remain small during time (see Supplementary Note 2).

Notably, wave turbulence consistently arises across a broad range of driving frequencies~\cite{navon_emergence_2016}, see for instance the time-evolution of the power-law exponents for $\omega_{\rm d} \in [1,5] \omega_x$ illustrated 
in Fig.~\ref{Fig:Momentum_exponents}(b).
Even though a quasi-steady state arises for all considered driving frequencies, $\omega_{\rm d}$ slightly affects the saturation value of the exponent at long evolution times.
In particular, at the high end of $\omega_{\rm d}$, $\omega_{\rm d}/(2\pi) \in [180,200]~\rm{Hz}$, the averaged exponent yields $\gamma =  2.70(24)$, while the mean value over all considered frequencies is $2.57(19)$ (dashed line in Fig.~\ref{Fig:Momentum_exponents}(b)).

\textbf{Impact of the initial state.} We subsequently investigate the role of the initial state on the turbulent response, by   
dynamically crossing the phase transition inversely, i.e., from a SF ($a_{\rm i}=98~a_0$) to a SS ($a_{\rm f}=91~a_0$).
At $t \gtrsim 300 ~ \rm{ms}$, significant density undulations 
arise which gradually develop into a wave turbulent cascade quantified by the power-law behavior of 
$\tilde{n}(\abs{\textbf{k}},t)$. 
The time-evolution of the associated exponents for
driving frequencies in the range $\omega_{\textbf{d}} \in [1,5]\omega_x$
is displayed in Fig.~\ref{Fig:Momentum_exponents}(c). 
They cluster around $2.64~(33)$ at $t > 1000~ \rm{ms}$ (dashed line in Fig.~\ref{Fig:Momentum_exponents}(c)),
namely 
at 
larger timescales 
compared to the opposite driving scenario [Fig.~\ref{Fig:Momentum_exponents}(b)]. 
However, restricting $\omega_{\rm d}/(2\pi)$ to $[130,200]~\rm{Hz}$ results in $2.83(26)$, which is closer to the prediction of weak wave turbulence in $d=3$. 
This timescale prolongation is reminiscent of a shaken non-dipolar SF in a box~\cite{navon_emergence_2016}, where the corresponding exponent converged towards $3.5$. 
Initially, $\gamma(t)$ heavily depends on the particular driving frequency, a response attributed to an initial pattern formation stage delaying the onset of turbulence and being strongly linked to $\omega_{\rm d}$ (see Supplementary Note 3).
Notice also that using  
a small driving amplitude $a_{\rm f}-a_{\rm i}$, where the dipolar gas remains within the same (SF or SS) phase,  
leads to weak density deformations 
without any evidence of wave turbulence (see Supplementary Note 3).

\begin{figure}[t!]
\centering
\includegraphics[width=\columnwidth]{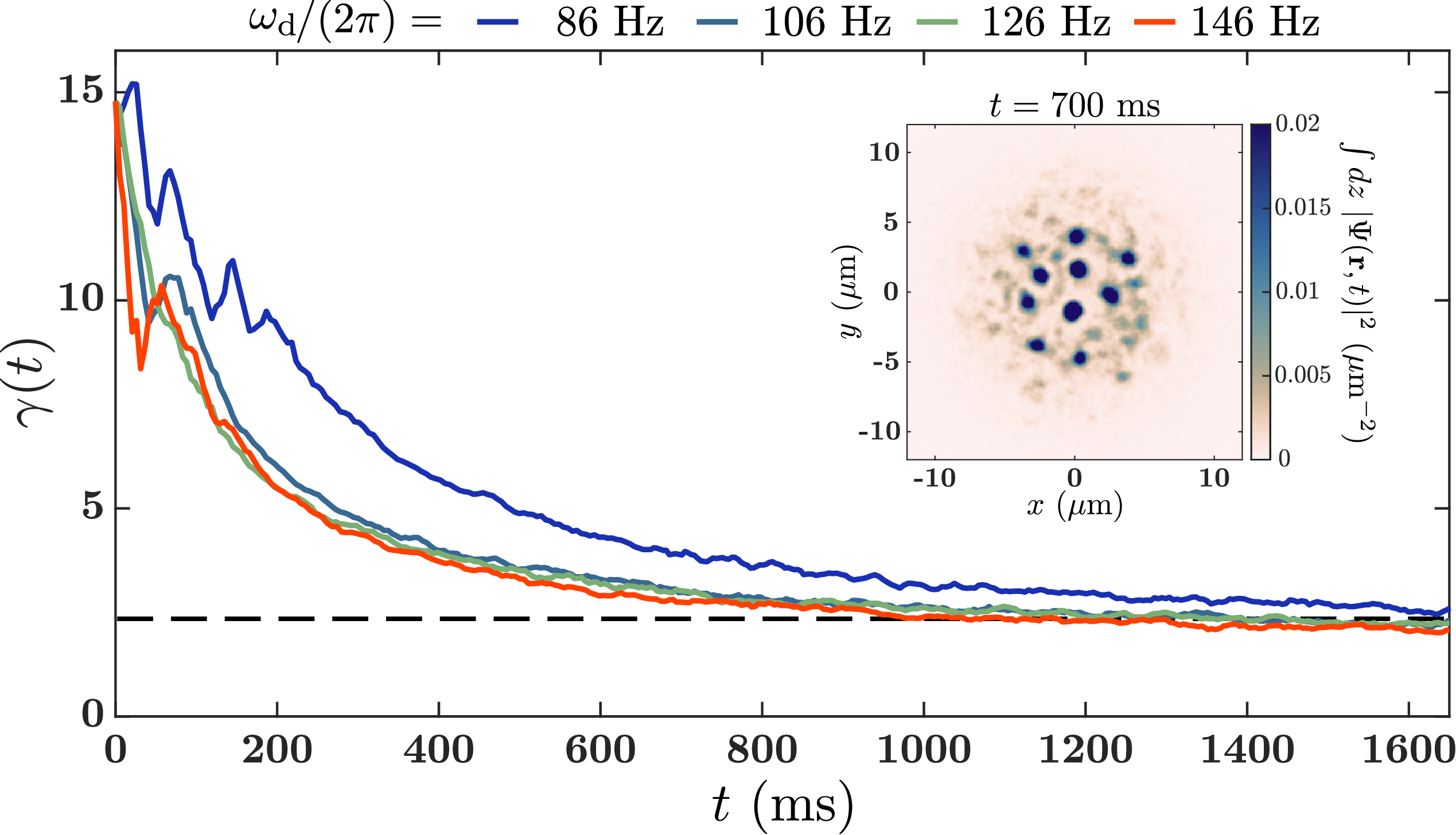}
\caption{
\textbf{Dynamical crossing of a SS to the isolated droplets regime leading to wave turbulence.} 
Time evolution of the scaling exponent
under continuous driving from the SS ($a_{\rm i}=89~a_0$) to droplets ($a_{\rm f}=82~a_0$) phase for different driving frequencies (see legend). 
Saturation of the exponent around $2.41$ (dashed line) 
indicates the approach to a quasi-steady state. 
The latter encompassing a SF background is visualized in the inset depicting 
$\int dz~ \abs{\Psi(\textbf{r},t)}^2$ 
pertaining to $\omega_{\rm d}=2\pi \times 126~\rm{Hz}$ at long evolution times.
}
\label{Fig:Exp_supersolid_droplet}
\end{figure}

For completeness, we next explore the properties of the nonequilibrium quasi-steady state when 
transitioning from a SS to a dipolar droplet state~\cite{ferrier-barbut_observation_2016}. 
We start from the 
SS configuration 
with  $a_{\rm i}=89~a_0$, see also Fig.~\ref{Fig:Density_profiles}(a), and deploy periodic modulation of the scattering length 
to $a_{\rm f}=82~a_0$. 
The latter gives rise to  
a localized array of isolated droplets 
in the quasi-2D ground state phase diagram~\cite{Baillie_droplet_2018}. 
However, in the course of the evolution a significant SF connection among the droplets persists, 
as shown in the inset of Fig.~\ref{Fig:Exp_supersolid_droplet} referring to $\omega_{\rm d}=2\pi \times 126~\rm{Hz}$. 
This SF background along with the high-density peaks---similarly to what has been observed in the SF-to-SS transition and vice versa---becomes substantially perturbed at long timescales. 
Crucially, the exponents $\gamma(t)$ obtained from the tails of $\tilde{n}(\abs{\textbf{k}},t)$ irrespectively of $\omega_{\rm d}$ [Fig.~\ref{Fig:Exp_supersolid_droplet}] attain  
the wave turbulence value 
$\gamma(t \gtrsim 900~\rm{ms}) \approx 2.41 \: (29)$, 
as indicated by the dashed line in Fig.~\ref{Fig:Exp_supersolid_droplet}.
These are the same timescales as in the SS-to-SF transition [Fig.~\ref{Fig:Momentum_exponents}(b)].
Such an observation raises the question: why does the SS state facilitate the transition to wave turbulence faster than a SF? 

\textbf {Cascade front dynamics.} \label{sec:energy}
The answer lies in the dynamics of the cascade front $\abs{\textbf{k}_{\rm cf}}$,  associated with the cusp point where $\tilde{n}(\abs{\textbf{k}},t)$ falls faster than $\abs{\textbf{k}}^{-\gamma}$ at large momenta~\cite{nazarenko_wave_2011}, 
and governing energy transport towards larger momenta in wave turbulence.  
As the cascade front propagates, it establishes the quasi-steady state in its wake. Within the 
weak wave turbulence framework~\cite{zakharov_kolmogorov_1992}, the growth of the cascade front follows the scaling relation $\abs{\textbf{k}_{\rm cf}} \sim t^{-\beta}$, with 
$\beta = -\frac{1}{2+d-\gamma}$ in $d$-dimensions.
It is derived from energy and particle conservation laws and holds universally, i.e., regardless of the system microscopic characteristics,  
as long as the dispersion relation of elementary excitations is quadratic~\cite{navon_synthetic_2019}.
For a dipolar condensate, 
the linear phononic contribution 
at low momenta is followed by the roton minimum 
at intermediate momenta, and 
eventually
the dispersion relation becomes 
quadratic~\cite{petter_probing_2019}.  
The roton minimum for the SS initial state considered herein corresponds to   $\abs{\textbf{k}_{\rm rot}} l_{\rm s} \simeq 0.1487$; see the vertical dash-dotted line in Fig.~\ref{Fig:Momentum_exponents}(a).
Therefore, it is anticipated that the energy growth rate to high momenta scales with the same exponent $\beta$ for both SS and SF initial states.

\begin{figure}[t!]
\centering
\includegraphics[width=\columnwidth]{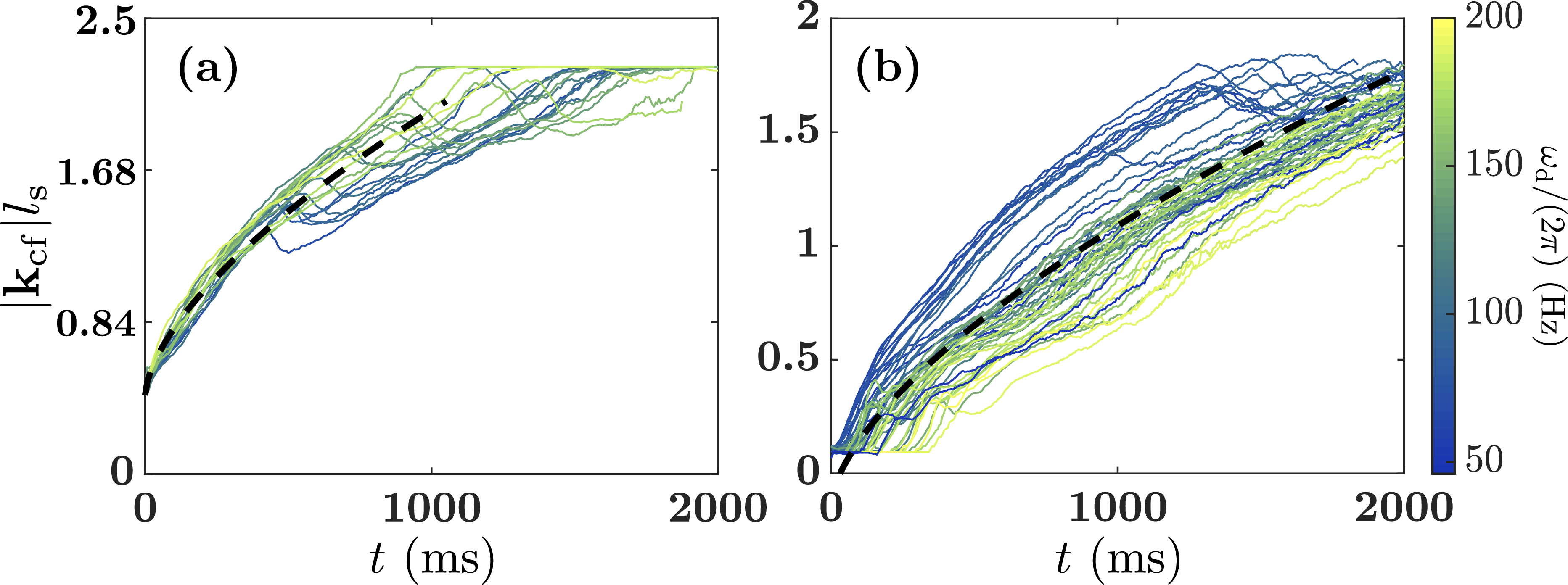}
\caption{
\textbf{Wave turbulence develops faster in a SS as compared to a SF.}
Cascade front $\abs{\textbf{k}_{\rm cf}} l_{\rm s}$ 
when transitioning from
(a) SS-to-SF [$a_{\rm i}~(a_{\rm f})=89~(98)~a_0$], and (b) vice versa [$a_{\rm i}~(a_{\rm f})=98~(91)~a_0$] 
for different driving frequencies, $\omega_{\rm d}$. The black dashed lines represent the fits $\abs{\textbf{k}_{\rm cf}} \sim t^{-\beta}$. 
The extended momentum distribution of the SS leads to a faster saturation of the cascade front compared to the SF.  
}
\label{Fig:wavefront}
\end{figure}

The cascade front is determined by the intersection of $\tilde{n}(\abs{\textbf{k}},t)(\abs{\textbf{k}} l_{\rm s})^{\gamma}$ and a threshold set at half of the nonequilibrium quasi-steady state plateau~\cite{Galka_emergence_2022}, see the horizontal dashed line in Fig.~\ref{Fig:Momentum_exponents}(a). 
Fitting $\abs{\textbf{k}_{\rm cf}}$ with $t^{-\beta}$ (dashed lines in Fig.~\ref{Fig:wavefront}), unveils the scaling exponent for the SS to SF phase crossing [Fig.~\ref{Fig:wavefront}(a)], $\beta=-0.63 \, (05)$, which  is indeed similar to the one extracted for 
the reverse crossing [Fig.~\ref{Fig:wavefront}(b)], that is $\beta=-0.64 \, (07)$. 
Moreover, the scaling 
is almost independent of $\omega_{\rm d}$, 
demonstrating once more the independence of the quasi-steady state on the driving. 
Additionally, the same scaling exponent $\beta$ appears in the transition of SSs to dipolar droplet lattice, further evincing the universality of wave turbulence.
In all cases a cascade front propagates also in the axial direction, evincing a direct energy cascade along $z$ (see Supplementary Note 1). 
However, no clear scaling law behavior is observed.

Although the cascade front 
exhibits a similar growth in both dynamical crossings of the SS-to-SF transition, it 
starts from a larger value when the system resides in a SS phase, see Fig.~\ref{Fig:wavefront}(a). 
This difference arises because the momentum distribution of the latter possesses a larger amplitude at large $\abs{\textbf{k}}$ compared to that of a SF, due to the high momentum peaks associated with the roton minimum~\cite{chomaz_dipolar_2022}.
Therefore, at a given time, $\abs{\textbf{k}_{\rm cf}}$ attains higher values in the case of a SS compared to a SF \footnote{The long-time behavior of the cascade front depends on the chosen threshold. Namely, smaller thresholds lead to saturation of $\abs{\textbf{k}_{\rm cf}}$ at larger momenta. However, the scaling exponent $\beta$ has a weak dependence on the threshold for both initial states.}.
As a consequence, the saturation of $\abs{\textbf{k}_{\rm cf}}$ to the largest available momentum scale in our setup occurs faster in the case of the SS-to-SF transition [Fig.~\ref{Fig:wavefront}(a)].

Apart from $\abs{\textbf{k}_{\rm cf}}$, the exponent $\beta$ dictates the evolution of the momentum distribution prior to the establishment of the quasi-steady state. In particular, within weak wave turbulence the scaling relation $\left( \frac{t}{t_0}  \right)^{-\gamma \beta} \tilde{n}(\abs{\textbf{k}},t) = \tilde{n} (\abs{\textbf{k}}(t/t_0)^{\beta},t_0)$ holds~\cite{Galka_emergence_2022}, 
where  
$t_0$ is an arbitrary timescale. 
Exploiting this relation reveals that the curves of $\tilde{n}(\abs{\textbf{k}},t) (\abs{\textbf{k}}l_{\rm s})^{\gamma}$ 
at long time instants converge towards a narrow band regardless of the initial state unveiling the self-similar dynamics of the turbulent cascade.

\textbf{Turbulence in the presence of three-body losses}.  \label{Sup:Losses}
It is known that the many-body self-bound SS and droplet states suffer from three-body recombination processes experimentally~\cite{bottcher2019transient} since they exhibit high densities. 
A question that arises is whether the relevant three-body loss mechanisms might impede the experimental detection of wave turbulence, given that it occurs over long evolution times.
To facilitate corresponding experimental endeavors, here, we demonstrate that the inclusion of three-body losses does not prevent the emergence of wave turbulence. 
We substantiate this argument by using the SS as an initial  state with $a_{\rm i}=89 ~ a_0$, and subsequently dynamically cross the SF boundary upon considering $a_{\rm f}=98 ~ a_0$ and a driving frequency $\omega_{\rm d} = 2\pi \times  127 ~ \rm{Hz}$. 
To take into account three-body losses, 
we consider an additional imaginary term in our extended Gross-Pitaevskii description emulating such lossy channels with the recombination rate
$L_3=1.2 \times 10^{-41}~\rm{m}^6/\rm{s}$, 
pertaining to $^{164}$Dy atoms~\cite{Wachtler_quantum_2016,casotti_observation_2024}, see Methods for details.

\begin{figure}[t!]
\centering
\includegraphics[width=1\columnwidth]{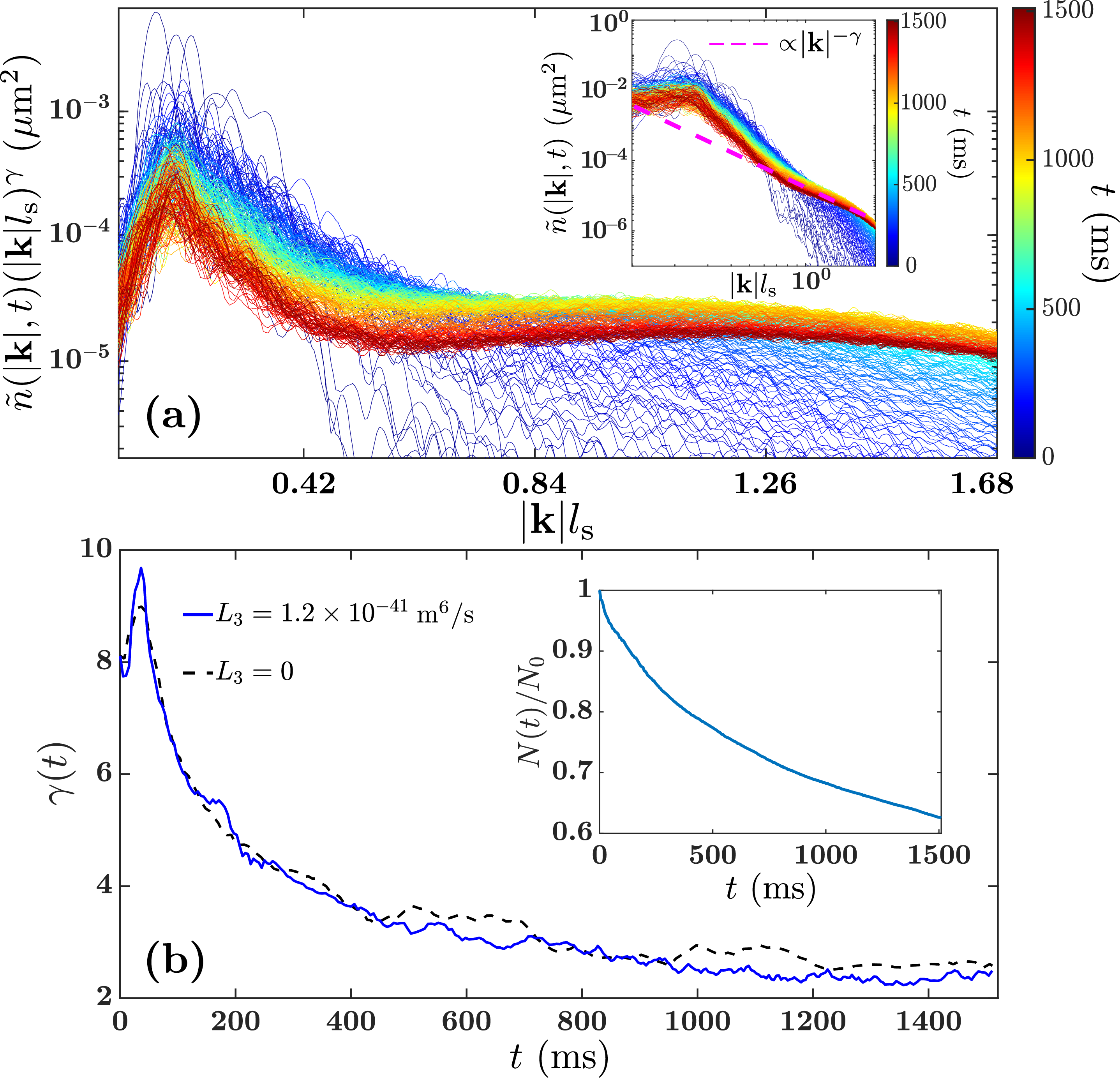}
\caption{
\textbf{Onset of wave turbulence in a driven SS in the presence of three-body recombination}. 
(a) Compensated spectrum with the inclusion of three-body losses for $a_{\rm i}=89~a_0$, $a_{\rm f}=98~a_0$ and $\omega_{\rm d}=2 \pi \times 127 ~{\rm Hz}$ as 
in Fig.~\ref{Fig:Momentum_exponents}(a). Saturation of the spectrum at $t>500~{\rm ms}$ takes place manifesting the approach to turbulent behavior. The inset presents the $\abs{\textbf{k}}^{-\gamma}$ power-law fitting of the momentum distribution in the range $\abs{\textbf{k}} l_{\rm s} \simeq 1$. (b) Comparison between the exponent dynamics at large momenta with finite and zero three-body recombination (see legend). As it can be seen, there are no appreciable deviations imprinted in the exponent between the finite $L_3$ and $L_3=0$ scenario, while in both cases the exponent saturates close to $\approx 2.51$. The inset presents the particle number during the time evolution, normalized to the initial number $N_0=8 \times 10^4$. Even for the $\sim 40 \%$ atom losses occurring at long evolution times a SS structure persists.}
\label{Fig:Supersolid_three_body}
\end{figure}

Despite the occurrence of three-body losses, the dipolar SS still clearly enters the wave turbulent regime at long evolution times. This is evident by the power law behavior at large momenta shown in the inset of Fig.~\ref{Fig:Supersolid_three_body}(a), with $\gamma = 2.51(05)$. 
In particular, a plateau appears in the compensated spectrum [Fig.~\ref{Fig:Supersolid_three_body}(a)] around $\abs{\textbf{k}} l_{\rm s} \simeq 1$ at $t>600 ~ \rm{ms}$, signaling the emergence of a nonequilibrium quasi-steady state.
A similar behavior of the compensated spectrum occurs also for other driving frequencies (not shown for brevity).
The fitted exponent [solid line in Fig.~\ref{Fig:Supersolid_three_body}(b)] lies very close to the one pertaining to $L_3=0$ (dashed line) during the dynamics, displaying negligible deviations at late evolution times.
Both of the exponents eventually saturate in the ballpark of $2.5$. 
For completeness, we remark that within the presented timescales the particle number drops to around $\simeq 60 \% $ of the initial atom  number $N_0=8 \times 10^4$, see the inset of Fig.~\ref{Fig:Supersolid_three_body}(b). 
This is also manifested by the smaller amplitude of the compensated momentum distribution [Fig.~\ref{Fig:Supersolid_three_body}(a)] in comparison to the respective one without three-body losses [Fig.~\ref{Fig:Momentum_exponents}(a)].
This behavior is particularly pronounced at large momenta, where $\tilde{n}(\abs{\textbf{k}},t) (\abs{\textbf{k}} l_{\rm s})^{\gamma}$ [Fig. \ref{Fig:Supersolid_three_body}(a)] displays a much steeper descend than without the $L_3$ coefficient.  
This prominent descend at large momenta facilitated by three-body losses is in line with experimental observations~\cite{Galka_emergence_2022}.
Moreover, the SS nature of the dynamical state is still preserved within  the timescales that the quasi-steady state is reached.

{\vspace{0.25cm}\noindent\large\textbf{Conclusions}} \label{sec:conclusions}

We 
theoretically explored the manifestation and properties of wave turbulence in driven dipolar gases featuring beyond mean-field corrections,  
while dynamically crossing the ensuing phase boundaries. 
The  behavior of the spectrum at long timescales is unveiled associated with a direct energy cascade and the emergence of a genuine nonequilibrium quasi-steady state. 
The estimated scaling exponents at large momenta signal the onset of wave turbulence. 
Interestingly, a similar power-law behavior and exponent appear in the presence of three-body losses which, however, enforce a more drastic decay at large momenta.
The inherent extended momentum distributions of exotic phases of matter such as SS or droplets expedite the generation of the ensuing nonequilibrium quasi-steady state. 
We have demonstrated the emergence of wave turbulence independently of the initial phase, the driving frequency and the crossed phase boundary.   

Our results open new avenues for the exploration of quantum turbulence and the universal features of nonequilibrium dynamics in long-range interacting platforms~\cite{chomaz_dipolar_2022}, even beyond dipolar gases, exhibiting, for instance, crystalline density order~\cite{Choi_turbulent_2002}. 
In this context, it would be valuable to explore how different exotic crystal 
arrangements~\cite{ripley_two-dimensional_2023,Hertkorn_pattern_2021} (beyond the hexagonal one) 
can be systematically used to engineer the onset of turbulence. 
A natural next step is to investigate the emergence of the inverse energy cascade~\cite{karailiev_observation_2024,Reeves_inverse_2013,Neely_characteristics_2013} in such long-range interacting setups with a particular focus on understanding the role of relevant beyond-mean-field phases of matter. 
Another interesting pathway is to explore non-conventional anisotropic turbulent aspects e.g. with the aid of a rotating magnetic field~\cite{Halder_control_2022,Mukherjee_stacks_2023}. 
Moreover, other dynamical protocols for generating wave turbulence would be worth pursuing, 
such as driven spatially modulated  external potentials~\cite{karailiev_observation_2024} 
reflecting the symmetry of the SS  configurations.
Finally, further studies are required to establish a solid  understanding of the impact of the dimensionality of the dipolar gas on the emergent turbulent cascades~\cite{Alexakis_quasi_2023,Danilov_quasi_2000}.

{\vspace{0.25cm}\noindent\large\textbf{Methods}}

The ground states and non-equilibrium dynamics are obtained by means of the imaginary and real time propagation respectively, employing the split-step Crank-Nicolson numerical scheme~\cite{crank_practical_1947,antoine_computational_2013}.
To emulate 
three-body losses in the dynamics of the 3D dipolar gas, we include the additional imaginary term $-\text{i} \hbar \frac{L_3}{2} \abs{\Psi}^4 \Psi$ to the right hand side of Eq.~\eqref{Eq:MF_equation}, with $L_3$ denoting the underlying experimentally relevant  three-body recombination rate for $^{164}$Dy atoms.
Let us note that despite the modulated scattering lengths, the recombination rate is taken to be time-independent. This is due to the fact that the background scattering length of $^{164}$Dy is roughly $92 ~ a_0$~\cite{Maier_broad_2015}. Therefore, relatively small scattering length variations are required to cross the SF-to-SS transition, translating to small changes in $L_3$.

{\vspace{0.25cm}\noindent\large\textbf{Data Availability}}
The data associated with this work are available from the corresponding author upon request.

{\vspace{0.25cm}\noindent\large\textbf{Code Availability}}
The code associated with this work are available from the corresponding author upon request.

{\vspace{0.25cm}\noindent\large\textbf{Author Competing Interests}}
The authors declare no competing interests.

{\vspace{0.25cm}\noindent\large\textbf{Author Contributions}}

G.A.B. performed the numerical simulations and carried out the associated analysis. K.M. benchmarked part of the simulations and developed numerical scripts for portions of the analysis. S.I.M. conceived the idea of the work, supervised and funded the project. All authors contributed to the interpretation of the results and the writing of the manuscript.

{\vspace{0.25cm}\noindent\large\textbf{Acknowledgements}}

 S. I. M acknowledges support from the Missouri University of Science and Technology, Department of Physics, in the framework of a Startup fund. 
Financial support by the Knut and Alice
Wallenberg Foundation  and the Swedish
Research Council are also acknowledged (K. M). 
S. I. M acknowledges extensive discussions with H. R.  Sadeghpour in the context of universal dynamics and supersolid character. 
K.M  gratefully acknowledges many discussions with Stephanie M. Reimann on the topic of supersolidity.
The authors acknowledge the anonymous referees for their insightful comments.

\putbib[Dipolar_Gases]

\end{bibunit}

\clearpage

\begin{bibunit}[apsrev4-1]
    

\onecolumngrid
\setcounter{equation}{0}
\setcounter{figure}{0}
\setcounter{section}{0}
\makeatletter
\renewcommand{\theequation}{S\arabic{equation}}
\renewcommand{\thefigure}{S\arabic{figure}}
\renewcommand{\bibnumfmt}[1]{[S#1]}
\renewcommand{\citenumfont}[1]{S#1}
\renewcommand{\thesection}{\arabic{section}}
\setcounter{page}1
\def\thepage{S\arabic{page}}

\begin{center}
	{\Large\bfseries Supplementary Material: Generation of wave turbulence in dipolar gases driven across their phase transitions \\ 
 }
\end{center}

\section{Supplementary Note 1: Dynamics in the axial direction} \label{Sec:Transverse}

\begin{figure}[h!]
\centering
\includegraphics[width=1\columnwidth]{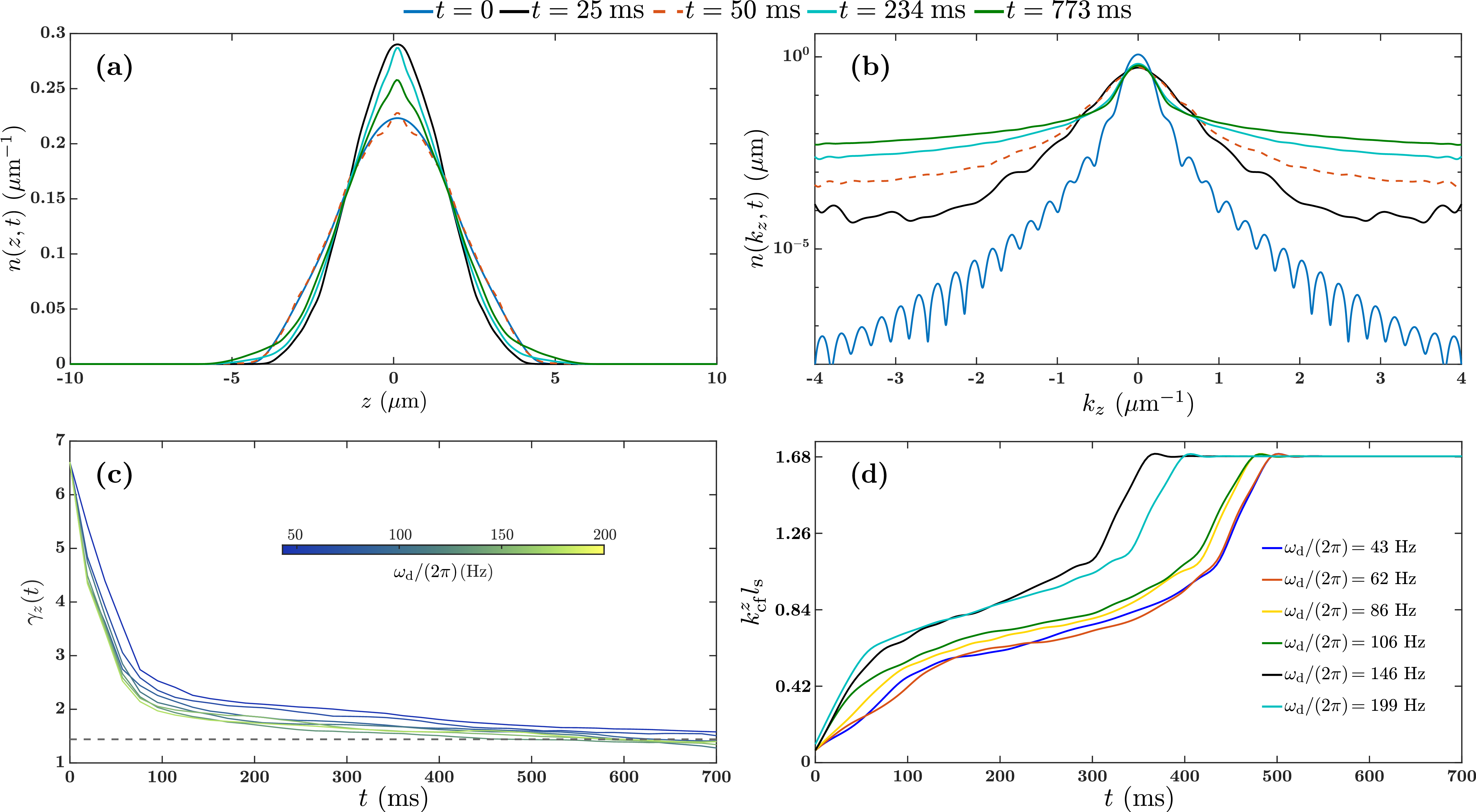}
\caption{(a) Integrated density snapshots (see legend) along the axial $z$-direction upon dynamically crossing the supersolid-to-superfluid transition in a periodic manner with $\omega_{\rm d}/(2\pi) = 127 ~ \rm{Hz}$. 
As it can be seen, the density dynamics in the $z$-direction is basically restricted to a weak amplitude collective motion. 
(b) The corresponding momentum distribution reveals relatively  enhanced 
tails, reflecting the presence of the small length scale density fluctuations [panel (a)] and signaling the transfer of energy in the $z$ direction. (c) Scaling exponents, $\gamma_z(t)$, quantified through the algebraic reduction of $n(k_z,t)$ at large momenta. Saturation occurs close to $1.4$ (dashed line) over a large range of driving frequencies (see legend). (d) Time-evolution of the cascade front, $k^z_{\rm{cf}}$,  capturing the energy transfer in the $z$ direction.}
\label{Fig:Transverse_profile}
\end{figure}

In the main text, the dynamics of the planar density is tracked, revealing the emergence of a non-equilibrium quasi-steady state at long evolution times.
Due to the considered trap  
aspect ratio ($\omega_z/\omega_r=3$) though, the dynamics cannot be considered as purely 2D, since the transverse excitation energy is smaller than the chemical potential of the initial state, i.e. $\hbar \omega_z < \mu$.
It is therefore rather a quasi-2D setup in the sense that the high-density droplet crystals are arranged in the plane, setting also a characteristic momentum scale, the roton momentum.
To appreciate the dynamics in the $z$ direction in the case of the supersolid-to-superfluid crossing, the integrated density $n(z,t) = \int dx dy ~ |\Psi(x,y,z,t)|^2$ is monitored, see Fig.~\ref{Fig:Transverse_profile}(a).
As time evolves, it becomes apparent that the density $n(z,t)$ features a relatively weak amplitude collective expansion and contraction dynamics, which becomes more prominent as the dipolar gas enters the turbulent regime.
This is due to the gradual melting of the supersolid crystals which are initially elongated along the axial direction [see also Fig. 1 in the main text].
The collective dynamics is, of course, less violent compared to the planar motion, precisely due to the quasi-2D setting.
Interestingly, the periodic driving causes localized density modulations around $z=0$, leading to hump-like structures at small length scales, see Fig.~\ref{Fig:Transverse_profile}(a) at $t=773~$ms.
This is further corroborated by the corresponding momentum distribution $n(k_z,t) = \int dk_x dk_y ~ |\mathcal{F}[\Psi](k_x,k_y,k_z,t)|^2$, with $\mathcal{F}[\Psi]$ denoting the Fourier transform of the 3D  wavefunction [Fig.~\ref{Fig:Transverse_profile}(b)].
As time progresses, 
the large $k_z$ momentum tails become significantly populated (as an imprint of the localized density humps in $n(z,t)$), evincing energy transfer processes in the axial direction as well.

To further understand this energy transfer, we inspect the high momentum tails of $n(k_z,t)$, which display an algebraic scaling at large momenta, $n(k_z \gg 1,t) \simeq k_z^{-\gamma_z(t)}$. The associated exponent, $\gamma_z(t)$, saturates at long evolution times [Fig.~\ref{Fig:Transverse_profile}(c)] close to $\gamma_z \simeq 1.4$ [see also the dashed line]. 
To compare the exponent with the one resulting from the planar dynamics, we employ the line-of-sight integrated distribution, $n_x(k_x,t) = \int dk_y ~ n(\textbf{k},t)$.
The corresponding exponents, $\gamma_x$, i.e. $n_x(k_x \gg 1,t) \simeq k_x^{-\gamma_x(t)}$, tend asymptotically to $\approx 1.57$. 
They are reduced as compared to the ones reported in the main text, $ \gamma \approx 2.57$, due to the integration~\cite{navon_emergence_2016}.
The discrepancy between $\gamma_x$ and $\gamma_z$ reveals that
the turbulent cascade develops differently in the plane compared to the z-direction.
Such a difference can be attributed to the 2D crystal structure of the dipolar gas in the supersolid phase, which results in anisotropic momentum distributions during the turbulent cascade.
However, the origin of the saturated $\gamma_z$ value remains an open question for future investigations, aiming to investigate the impact of different dimensionalities and their crossover on the observed scaling exponents. Similarly to the planar dynamics, a cascade front $k^z_{\rm{cf}}$ in the axial direction can be defined, stemming from the compensated spectra $n(k_z,t) (k_z l_{\rm s})^{\gamma_z}$, where $\gamma_z$ is the saturated exponent. The cascade front [Fig.~\ref{Fig:Transverse_profile}(d)] clearly indicates a direct energy cascade. However, the temporal behavior of $k^z_{\rm{cf}}$ deviates significantly from the power law behavior observed in the case of the  planar dynamics.

\section{Supplementary Note 2: Interplay of energy contributions} \label{Sec:Energy}

\begin{figure}[t!]
\centering
\includegraphics[width=1\columnwidth]{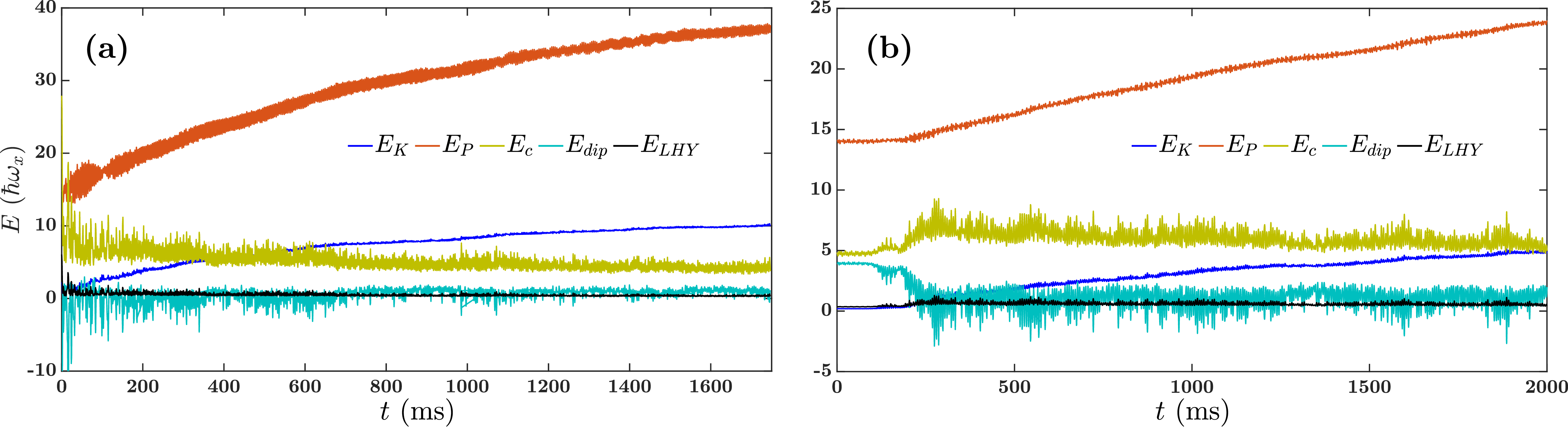}
\caption{Individual energy contributions (see legend and text) in the case of the (a) supersolid-to-superfluid and (b) superfluid-to-supersolid dynamical crossings.  
The kinetic energy increases over time in both cases, 
evincing the development of wave turbulence. 
All interaction contributions show a decreasing tendency in the course of the evolution, while the potential energy dominates due to spreading of the density distribution in the plane. 
The used driving frequencies correspond to (a) $\omega_{\rm d}/(2\pi) = 127 ~ \rm{Hz}$ and (b) $\omega_{\rm d}/(2\pi) = 85 ~ \rm{Hz}$.}
\label{Fig:Energy_contributions}
\end{figure}

The onset of wave turbulence reported in the main text for dipolar gases signifies the transfer of energy from large to small length scales.
This is clearly illustrated by the extended high momentum tails of the momentum distribution presented in Fig. 2(a) in the main text.
Such an energy transfer occurs primarily due to single particle excitations, and thus the kinetic energy, $E_K$, continuously increases.
This mechanism can be readily testified by inspecting 
Fig.~\ref{Fig:Energy_contributions} depicting the evolution of the participating energy contributions for both the supersolid-to-superfluid and the superfluid-to-supersolid dynamical transitions using the periodically modulated scattering length protocol  discussed in the main text.  
Notably, $E_K$ starts to increase from early evolution times when the system is initialized in the supersolid phase [Fig.~\ref{Fig:Energy_contributions}(a)], a behavior that is attributed to the faster onset of wave turbulence displayed by supersolids.

Apart from the kinetic energy, the potential term, $E_p$, shows an increasing tendency and in fact it is the most dominant one.
In the turbulence regime, the dipolar gas becomes substantially excited and expanded across the plane resulting in a prevalent 
$E_p$ contribution.
This is accompanied by a decrease in the energy due to dipolar interactions, $E_{dip}$.
Moreover, the energy associated to the beyond mean-field correction term, $E_{LHY}$, is negligible at late evolution times. 
Its contribution is significant only at the initial time instants for the supersolid-to-superfluid crossing [Fig.~\ref{Fig:Energy_contributions}(a)]. 
This occurs since the initial state (supersolid) exhibits high density crystal peaks [see also Fig. 1(a) in the main text].
Subsequently, the crystals melt and so $E_{LHY}$ rapidly diminishes.
Furthermore, the energy attributed to contact interactions, $E_c$, features a small decreasing tendency, especially at large evolution times ($t > 800 ~ \rm{ms}$ in Fig.~\ref{Fig:Energy_contributions}), since atoms in the turbulent regime tend to slightly spread apart.

\section{Supplementary Note 3: Absence of turbulence for driving within the same phase} \label{Sup:Small_amplitude}

\begin{figure}[h!]
\centering
\includegraphics[width=0.8\columnwidth]{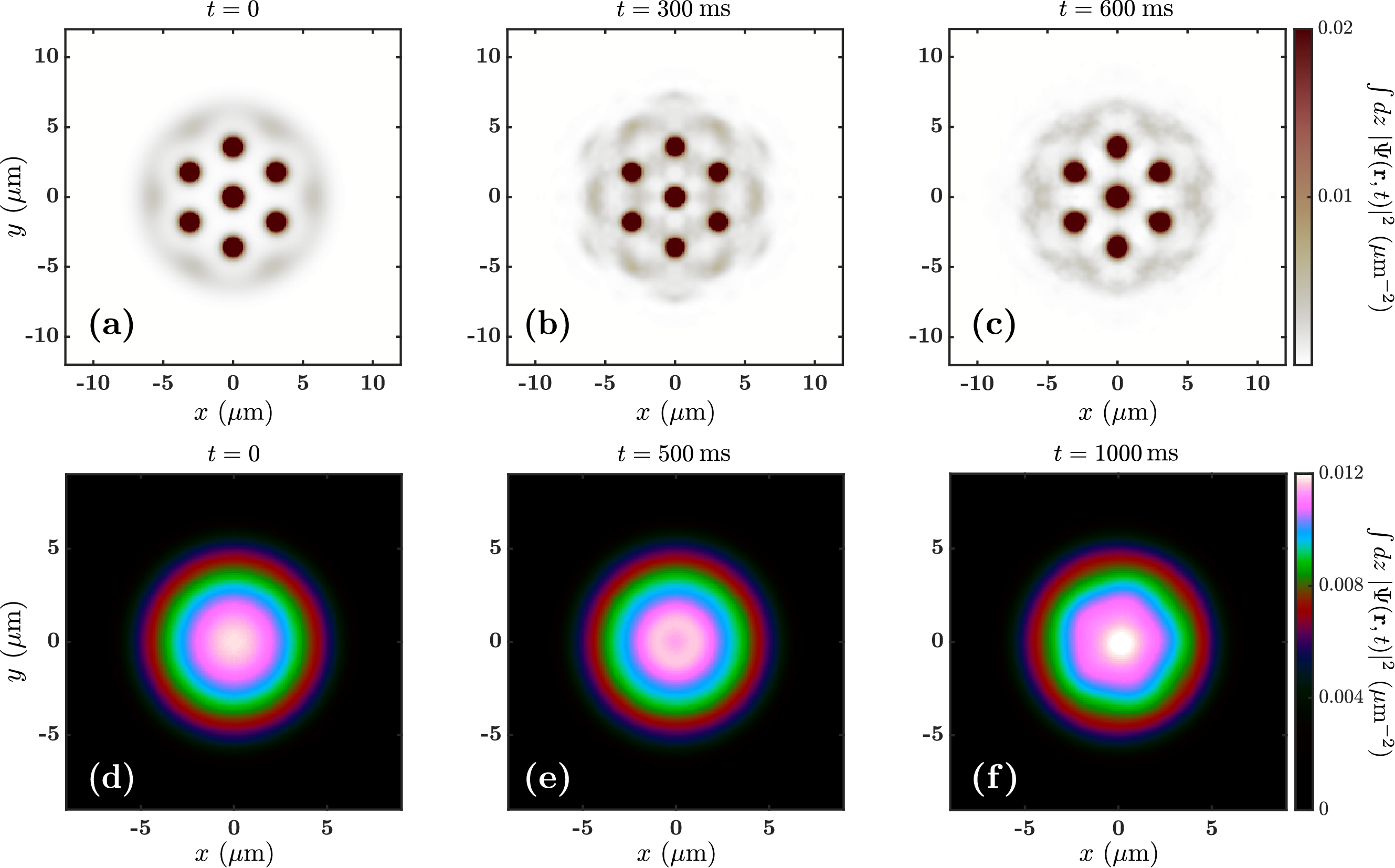}
\caption{Absence of wave turbulence for a dipolar gas driven within the same phase. Density snapshots (integrated along the axial direction) across the $x$-$y$ plane for an initial (a)-(c) SS with $a_{\rm i}=89~a_0$ and (d)-(f) a SF at  $a_{\rm i}=98~a_0$. 
Evidently, the SS features a distorted background while the SF performs a collective motion. 
The driving amplitude, $\abs{a_{\rm f}-a_{\rm i}} = 2 ~a_0$, is relatively small in both cases such that the underlying ground state phase boundary is not dynamically crossed. 
The used driving frequencies are $\omega_{\rm d} = 2\pi \times  127 ~ \rm{Hz}$ and $\omega_{\rm d} = 2\pi \times   85 ~ \rm{Hz}$ for the SS and SF respectively.}
\label{Fig:Small_driving_amplitude}
\end{figure}

In the main text, it is argued that, regardless of the initial state (SF, SS, or droplet), when the dipolar gas is periodically driven across one of the resulting phase transitions, it exhibits wave turbulence in the later stages of the evolution. Here, we demonstrate that employing sufficiently small driving amplitudes, $a_{\rm f}-a_{\rm i}$, which ensure that the dipolar gas remains in the same phase dynamically, prevents the manifestation of a turbulent response, irrespective of the initial state.
To substantiate our argument we present the nonequilibrium dynamics when the dipolar BEC is initiated either in the SF or the SS state and it is subsequently under the influence of a sinusoidally modulated 3D scattering length with an amplitude that does not exceed any relevant phase boundary.  
The density snapshots of the driven dipolar gas are presented in Fig.~\ref{Fig:Small_driving_amplitude}.
Starting from a SS state characterized by $a_{\rm i} = 89~a_0$ we observe that, upon driving with  $a_{\rm f}=91~a_0$ and $\omega_{\rm d}= 2\pi \times   127 ~ \rm{Hz}$, only relatively small amplitude density undulations 
develop on top of the original hexagonal structure distorting the background SF but not the droplet arrangement, see the integrated density profiles depicted in Fig.~\ref{Fig:Small_driving_amplitude}(a)-(c).   
This response remains the same for longer timescales and holds for different driving frequencies that we have checked.  

Turning to a driven dipolar SF with $a_{\rm i}=98~a_0$, $a_{\rm f}=96~a_0$ and $\omega_{\rm d} = 2\pi \times   85 ~ \rm{Hz}$, illustrated in Fig.~\ref{Fig:Small_driving_amplitude}(d)-(f), it can be seen that collective motion of the background occurs in the form of an isotropic breathing mode. 
This is accompanied by signatures of developing bulk structures such as the pentagon pattern appearing in Fig.~\ref{Fig:Small_driving_amplitude}(f). 
The spontaneous generation of such patterns is another interesting venue for future research in terms of exploiting parametric instabilities~\cite{Kwon_stars,liebster2023emergence}.  
Let us finally note that when the driving amplitude $a_f-a_i$ is small but sufficient to cross the phase boundary, wave turbulence characteristics develop at long evolution times (not shown).
The onset of turbulence is slower compared to the case where the driving amplitude is large.

\section{Supplementary Note 4: Turbulence in the presence of dissipation} \label{Sup:damping}

As demonstrated in the main text, energy is transported to smaller length scales by means of nonlinear wave interference  
resulting in wave turbulence. Additionally, we have considered a fully condensed gas without any damping channels. In the following, we study the emergent non-equilibrium dynamics by including a phenomenological damping term, $\kappa$, to the extended Gross-Pitaevskii equation~\cite{Choi_phenomenological_1998,pitaevskii_phenomenological_1959,bland_two-dimensional_2022,Linscott_thermally_2014},
\begin{gather}
(\text{i}-\kappa) \hbar \partial_t \Psi(\textbf{r},t) = \Bigg [ - \frac{\hbar^2}{2m} \nabla^2 +V(\textbf{r})  + \frac{4\pi \hbar^2 a}{m} \abs{\Psi(\textbf{r},t)}^2  
+ \int d\textbf{r}'~ U_{\rm dd}(\textbf{r} -\textbf{r}')\abs{\Psi(\textbf{r}',t)}^2  
+  f (\epsilon_{\rm dd}) \abs{\Psi(\textbf{r},t)}^3 \Bigg ] \Psi(\textbf{r},t).
\label{Eq:MF_damping}
\end{gather}
The above model accounts for collisions of condensed atoms with those occupying the thermal fraction.
Such processes lead to the damping of excitations in the condensed portion without any accompanied particle loss.
The damping factor $\kappa$ usually depends on temperature, and is inversely proportional to the relaxation time of the dipolar BEC~\cite{proukakis_self_2024}.
The significance of $\kappa$ becomes apparent when the above equation is expressed in the hydrodynamical formalism, where $\hbar \kappa / (2m)$ plays the role of viscosity, dubbed the kinematic quantum viscosity~\cite{Bradley_2012_spectra}.

\begin{figure}[t!]
\centering
\includegraphics[width=1\columnwidth]{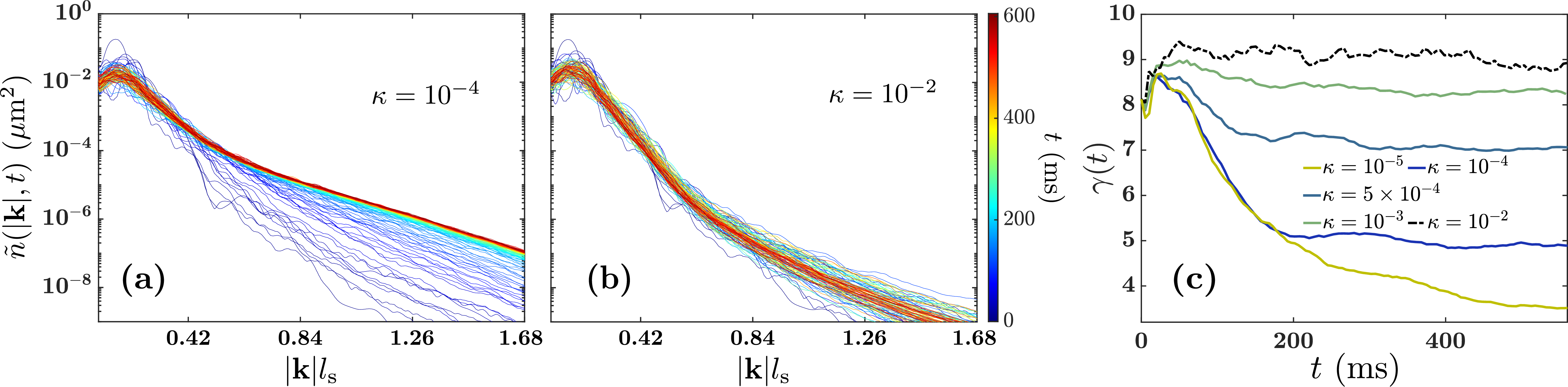}
\caption{Emergence of quasi-steady state at long evolution times in the presence of dissipation. Dynamics of the momentum distributions of an initial SS ($a_{\rm{i}} = 89 ~ a_0$) undergoing through the SS-to-SF transition with $a_{\rm f} = 98 ~ a_0$, $\omega_{\rm d} = 2\pi \times 127 ~ \rm{Hz}$ and phenomenological damping (a) $\kappa = 10^{-4}$ and (b) $\kappa = 10^{-2}$ respectively. (c) Time evolution of the exponents at large momenta for different phenomenological damping terms (see legend). A large $\kappa$ facilitates the suppression of the large momentum tails, and hence it leads to significantly elevated exponents at late time instants.}
\label{Fig:Supersolid_damping}
\end{figure}

To develop an understanding on how dissipation affects the onset of turbulence for the SS-to-SF dynamical crossing, we consider $\kappa$ in the range $[10^{-5}, 10^{-2}]$. 
The influence of damping can be readily seen by contrasting the momentum distributions in Fig.~\ref{Fig:Supersolid_damping}(a),(b). 
A larger $\kappa$ leads to the decay of high momentum excitations of the system that would otherwise be responsible for carrying energy to small length scales in the direct turbulent cascade.
Instead, the high momentum tail of $\tilde{n}(\abs{\textbf{k}},t)$ is highly suppressed, and even at large evolution times, the distribution remains close to the initial one.
Such an equilibration process is observed also in the time evolution of the extracted exponent at large momenta  [Fig.~\ref{Fig:Supersolid_damping}(c)]. 
For $\kappa = 10^{-5}$, which represents adequately small dissipation, the exponent quickly decreases 
close to the value obtained in the absence of  dissipation. 
For a larger $\kappa = 10^{-4}$, $\gamma(t)$ still saturates, but to a plateau whose value is larger than what is anticipated for weak wave turbulence.
As the damping term increases further, $\gamma(t)$ fluctuates close to its initial value.

\section{Supplementary Note 5: Simulation details} \label{Sup:Numerics}

To numerically tackle the ground state properties and the nonequilibrium dynamics of the considered quasi-2D dipolar gas we simulate the extended Gross-Pitaevskii equation~(1) provided in the main text. For numerical convenience it is casted in a dimensionless form. Specifically, space and time scales are expressed in units of the harmonic oscillator length  $a_{\rm ho}=\sqrt{\hbar / (m \omega_x)}$ and $\omega_x^{-1}$ respectively. Moreover, the 3D wave function is rescaled according to $\Psi(\textbf{r},t) = \sqrt{N/a^3_{\rm ho}} \Psi'(\textbf{r}',t')$, where the prime quantities refer to dimensionless ones. Note that this rescaling implies that the $\Psi'$ wave function is normalized to unity, i.e. $\int d\textbf{r}' \: \abs{\Psi'(\textbf{r}',t')}^2 =1$. 
Additionally, the three-dimensional space is discretized in a uniform cubic grid of $512 \times 512 \times 128$ points along the $x$, $y$ and $z$ directions respectively with 
spatial discretization $\delta x=\delta y=0.08 a_{\rm ho}$ in the plane and 
$\delta z=0.2a_{\rm ho}$ in the axial direction. 
The time step of the numerical integrator is set to $\delta t=10^{-4} / \omega_x$ 
for both the imaginary and real time propagation of Eq.~(1) in the main text.
These are carried out by means of the split-step Crank-Nicolson numerical scheme~\cite{crank_practical_1947,antoine_computational_2013}.
In order to deal with the divergent behavior of the integrand present in the dipolar interaction integral [Eq.~(1) in the main text], we utilize the convolution theorem~\cite{arfken_mathematical_1972,Goral_ground_2002}. In this way, the interaction integral is expressed as the inverse Fourier transform of the product of Fourier transforms of the dipolar interaction and wave function respectively. 
Therefore, the integral in Eq.~(1) of  the main text becomes numerically stable, without the need of any regularization scale.

To numerically obtain the exotic ground states of the dipolar gas, namely the SF, SS and droplet phases, the following initial ansatz is utilized 
\begin{equation}
\Psi'(x',y',z',t=0) = \mathcal{A} \Bigg[  \cos \left(  \frac{2\pi x'}{3} \right)  + \cos \left(  \frac{\pi x'}{3} + \frac{\pi y'}{\sqrt{3}}  \right) + \cos \left(  -\frac{\pi x'}{3} + \frac{\pi y'}{\sqrt{3}}  \right)  \Bigg ]^4 \text{e}^{-0.01 [  (x')^2 + (y')^2  ] -(\omega_z / \omega_x)  (z')^2/2 },
\label{Eq:Initial_ansatz}
\end{equation}   
with $\mathcal{A}$ being a normalization constant. 
This ansatz is particularly tailored for capturing the six-fold symmetry of the cylindrically symmetric SS configurations~\cite{Baillie_droplet_2018}.
During every step of the imaginary time propagation, the dimensionless wave function is rescaled, $\Psi' \to \Psi' /  \norm{\Psi'}$, where $\norm{\Psi'} \equiv \int d\textbf{r}' ~ \abs{\Psi'}^2$ denotes the norm, ensuring that the normalization is preserved.
The ground state is identified by requiring that the relative difference between the wave functions in two consecutive time steps of the imaginary time propagation is lower than $10^{-4}$ and the corresponding energy deviations do not exceed  $10^{-8}$.   
This state is subsequently employed for the real time dynamics of Eq.~(1) in the main text. 
Our spatial and time ($\delta t$) discretizations are chosen such that $(\delta t)^2 < \delta x\delta y$ holds and as a consequence the real time dynamics is accurately simulated. Namely, the total particle number and energy are numerically conserved within the order of $10^{-6}$ in the course of the evolution. 

\begin{figure*}[t!]
\centering
\includegraphics[width=1\columnwidth]{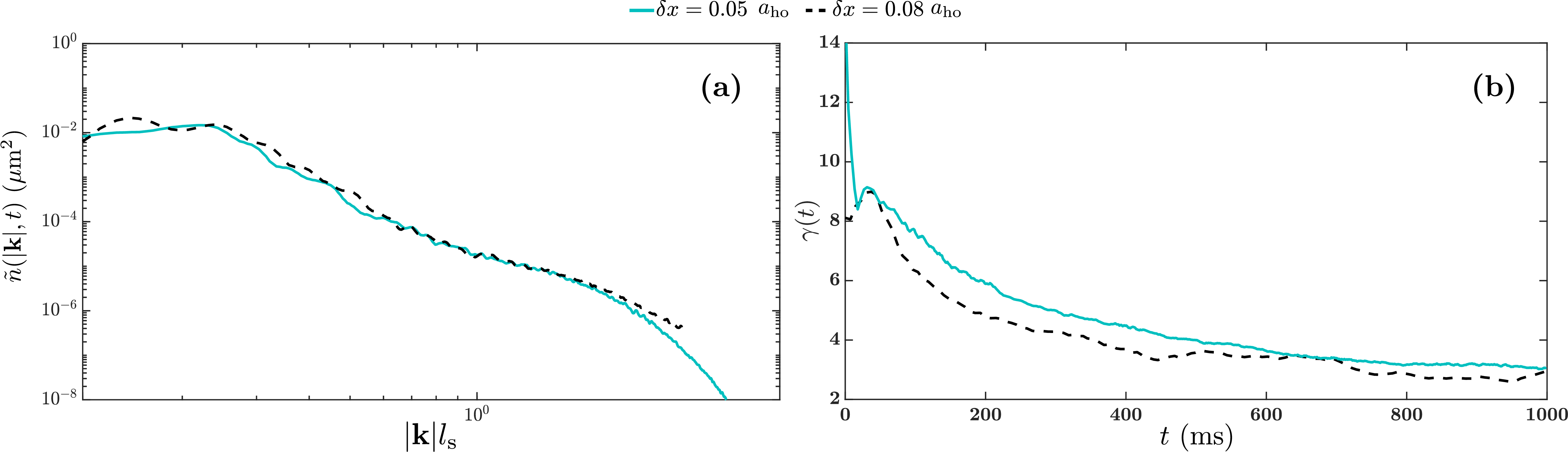}
\caption{Comparison of (a) the azimuthally averaged momentum distribution at $t=1000 ~ \rm{ms}$  and (b) the scaling exponent dynamics for 
different spatial discretizations (see legend) of the numerical simulations.  
Here, the supersolid-to-superfluid transition is dynamically crossed with $a_{\rm i} = 89 ~ a_0$, $a_{\rm f} = 98 ~ a_0$ and $\omega_{\rm d}/(2\pi) = 127 ~ \rm{Hz}$. 
}
\label{Fig:Fine_grid}
\end{figure*}

To further support the persistence of the turbulent characteristics discussed in the main text, we have performed simulations 
for smaller spatial discretizations, $\delta x =\delta y = 0.05 ~ a_{\rm ho}$ [Fig.~\ref{Fig:Fine_grid}].
As such, larger momenta can now be accessed compared to the case with  $\delta x = \delta y = 0.08 ~ a_{\rm ho}$ [compare the two lines in Fig.~\ref{Fig:Fine_grid}(a)] employed in the main text. 
The angularly averaged momentum distribution remains almost un-affected as can be seen in Fig.~\ref{Fig:Fine_grid}(a), referring to the supersolid-to-superfluid transition, at least for the momenta that can be accessed with larger spatial discretizations. 
Importantly, the exponent $\gamma(t)$ remains the same at long evolution times, saturating at the same value [Fig.~\ref{Fig:Fine_grid}(b)].
This value can now be extracted by fitting to a larger interval in momentum space, i.e. $\abs{\textbf{k}} l_{\rm s} \in [0.6,1.9]$, due to the access to larger $\abs{\textbf{k}}$ compared with the case pertaining to $\delta x  = \delta y= 0.08 ~ a_{\rm ho}$.
Discrepancies between the exponents stemming from the two different spatial discretizations occur mostly during the early evolution times.
This is due to the suppressed high momentum tails (large $\gamma(t)$), which are sensitive to the grid spacing. Similar conclusions can be drawn from the superfluid-to-supersolid phase transition. Summarizing, it is possible to infer the robustness of the observed dynamical response for different spatial discretizations.

\putbib[Supplement_bib]

\end{bibunit}

\end{document}